\documentclass[english]{elsarticle}
\usepackage[T1]{fontenc}
\usepackage[latin9]{inputenc}
\usepackage{textcomp}
\usepackage{graphicx}
\usepackage{babel}
\usepackage{todonotes}
\usepackage{hyperref}
\biboptions{sort&compress}
\begin{document}

\begin{frontmatter}{}

\title{Manipulating low-dimensional materials down to the level of single atoms with electron irradiation}

\author{Toma Susi}
\ead{toma.susi@univie.ac.at}
\author{Jannik C. Meyer}
\ead{jannik.meyer@univie.ac.at}
\author{Jani Kotakoski}
\ead{jani.kotakoski@univie.ac.at}
\address{University of Vienna, Faculty of Physics, Boltzmanngasse 5, 1090
Vienna, Austria}

\begin{abstract}
Recent advances in scanning transmission electron microscopy (STEM) instrumentation have made it possible to focus electron beams with sub-atomic precision and to identify the chemical structure of materials at the level of individual atoms. Here we discuss the dynamics that are observed in the structure of low-dimensional materials under electron irradiation, and the potential use of electron beams for single-atom manipulation. As a demonstration of the latter capability, we show how momentum transfer from the electrons of a 60-keV {\AA }ngstr{\"o}m-sized STEM probe can be used to move silicon atoms embedded in the graphene lattice with atomic precision. 
\end{abstract}
\end{frontmatter}{}

\section{Introduction}
Transmission electron microscopes (TEMs) and scanning TEMs (STEMs) can today be used to analyze matter at the level of single atoms. Their spatial resolution was held back for several decades by the pernicious problem of electron optic aberrations~\cite{Hawkes_aberrationsReview_2009}, finally solved by a concentrated effort in instrumentation development~\cite{Krivanek_firstCs_1997,Haider1998,Krivanek1999,Nellist_aberration_2004}. Despite their impressive resolution, until only a few years ago, microscopes were so sensitive that (in the scanning variety) focusing the electron beam on an individual atom for extended periods of time remained very hard, if not impossible. Moreover, the residual vacuum pressure in the instrument column meant that chemical etching of sensitive samples remained an issue over extended imaging times (see, for example, Refs.~\cite{Molhave2007,meyer_accurate_2012}). This situation changed with the advent of the Nion UltraSTEM that has an exceptionally stable sample stage and near-ultra high vacuum at the sample~\cite{krivanek_electron_2008}.

Due to the developments in electron microscopy~\cite{Haider1998,Krivanek1999,Erni2010b,krivanek_Z-contrast_2010,Dellby_40kV,krivanek_electron_2008}, several previously impossible feats are now possible, such as picometer-level structural mapping and strain analysis~\cite{Jia2008,Yankovich2014,Kimoto2010,Warner2011}, obtaining insights into the three-dimensional structure of small particles~\cite{VanAert2011a,BarSadan2008,Chen2013}, depth sectioning~\cite{Borisevich2006,Xin2008}, atomic resolution spectroscopic imaging~\cite{Muller2008,Pennycook2009, suenaga_nature_2010,Ramasse_Si_2013,Susi172DM}, measuring isotope concentrations at the nanometer scale~\cite{susi_isotope_2016} and the observation of dynamics at the level of single atoms~\cite{Batson2002,Varela2004a,Girit2009,Suenaga2009,Lee2013a,kotakoski2014imaging,susi_siliconcarbon_2014,lin_pyridinicN_2015}. Moreover, when studying two-dimensional (2D) materials such as graphene, hexagonal boron nitride, transition metal dichalcogenides, and various others~\cite{Nicolosi2013}, an {\AA }ngstr{\"o}m-sized electron beam can essentially be placed on each individual atom. This raises the intriguing question whether the atom-sized probe can in addition to analysis also be used for controlled atomic-level manipulation of low-dimensional materials.

Electron beams have long been an important tool for the shaping of matter on small scales. Due to their short wavelength and ease of control and focusing with electromagnetic lenses, electrons outperform any other type of radiation when it comes to reaching the smallest dimensions, rivaled only by ion beams~\cite{Jesse2016}. Probably the most widely used electron-based nano-fabrication application is so-called electron beam (e-beam) lithography, where the beams are used to define structures in a radiation-sensitive resist that are transferred to an underlying substrate via a chemical process~\cite{Ito2000}. However, the spatial dimensions of structures made by e-beam lithography are not limited by the size of the focused electron beam, but rather by the spread of the dose within the resist layer and other secondary processes. Other e-beam-based approaches are the direct cutting of matter with electron beams~\cite{Fischbein2007,Song2011,Xu2013,He2015}, the beam-activated etching of matter~\cite{Randolph2006,Yuzvinsky2005,Spinney2009,Thiele2013}, and the deposition of material from precursors under electron irradiation~\cite{VanDorp2014,VanDorp2005,Meyer2008b,Arnold2014}. Indeed, these resist-free approaches provide a versatile toolbox for structuring low-dimensional materials at the few-nanometer scale, only one order of magnitude larger than atomic dimensions.

In terms of the required doses, cross sections and threshold energies for such processes~\cite{Zobelli_KO_2007}, much depends on the specific material and chosen acceleration voltage. In metals, graphene in particular, radiolysis and ionization are in practice irrelevant due to the very fast valence and core hole recombination times (orders of magnitude faster than the time between electron impacts~\cite{susi_isotope_2016}). In semiconducting and insulating low-dimensional materials, these processes do play a role~\cite{Kotakoski_hBN_2010}, but are challenging to accurately quantify and disentangle from knock-on damage (KO)~\cite{Algara-Siller_MoS2gra_2013}. Furthermore, the local chemical environment greatly affects the bond strengths and thus also the interaction cross sections (e.g. weaker bound atoms neighboring an impurity or lattice edge). However, to give an idea of the magnitude of the relevant values, we consider in Table~\ref{tab1} several interesting processes in graphene: knock-on damage in the bulk and at the edge, a bond rotation in defective graphene, and knock-on damage and bond inversion processes of a C neighbor to a three-coordinated Si impurity.

\begin{table}
\begin{center}
       \caption{Typical values for the acceleration voltage $U$, order of magnitude of the characteristic dose $D_e$, cross section $\sigma$, and carbon displacement energy $T_d$ for irradiation-induced processes in graphene. \textit{KO} denotes knock-on damage, \textit{rota. @ DV} a bond rotation at a divacancy, \textit{ZZ} and \textit{AC} respectively graphene zigzag and armchair edges, \textit{C$_{\mathrm{Si}}$} a C neighbor to a Si impurity, \textit{exp.} experimental value, and \textit{DFTB} density-functional-based tight-binding simulated value. (Values in parentheses are estimated from the DFT thresholds for room temperature based on a vibration model using the phonon dispersion of pristine graphene~\cite{susi_isotope_2016}).\label{tab1}}
  \begin{tabular}{ c | c | c | c | c | c } \hline\hline
    Process & $U$ (kV) &$D_e$ (e$^-$/\AA$^2$) & $\sigma$ (barn) & $T_d$ (eV) & Ref. \\
       \hline
    KO @ bulk & 100 & 10$^9$ & 0.33 & 21.14 (exp.) & \cite{susi_isotope_2016} \\
    rota. @ DV & 80 & 10$^5$ & 2550 & 21-22 (DFTB) & \cite{kotakoski_stone-walestype_2011} \\
    KO @ ZZ & 60 & (10$^8$) & (2.8) & 12.0 (DFT) & \cite{Kotakoski_edges_2012} \\
    KO @ AC & 60 & (10$^{19}$) & (10$^{-11}$) & 19.0 (DFT) & \cite{Kotakoski_edges_2012} \\
    KO @ C$_{\mathrm{Si}}$ & 60 & 10$^{10}$ & 0.08 & 16.6 (exp.) & \cite{susi_siliconcarbon_2014} \\
    flip @ C$_{\mathrm{Si}}$ & 60 & 10$^8$ & 0.61 & 14.4 (exp.) & \cite{susi_siliconcarbon_2014} \\ 
    \hline
  \end{tabular}
\end{center}
\end{table}

In what follows, we will briefly review the controlled fabrication of structures on a few-nm scale as well as the beam-induced dynamics observed at the level of single atoms in low-dimensional materials. We will then show initial results that demonstrate a controlled (albeit limited) capability to move individual silicon atoms within the graphene lattice by several unit cells in a desired direction.

\section{Controlled structuring of 2D materials}

Miniaturization has been one of the main forces driving research into graphene and other low-dimensional materials, especially for electronic devices. Besides the use of such inherently small building blocks, creating them and manipulating their structure bring along a demand for a high degree of control and resolution. For example, the absence of an electronic band gap in graphene prevents its use as a transistor element; however, when it is shaped into a narrow ribbon, a gap is opened due to electron confinement (and possibly edge effects)~\cite{Stampfer2009}. Among the cornucopia of methods to form such ribbons (see, e.g., Ref.~\cite{Ma2013} for a review), e-beam-induced etching~\cite{Thiele2013} and the direct cutting of free-standing membranes~\cite{Song2011,Xu2013,MasihDas2016} fall into the category of resistless e-beam-based methods. A major drawback for electron- and ion-beam based cutting is the formation of defects, amorphous areas and contamination on the boundary of the removed sections, which can, however, be mitigated to some extent by keeping the sample at elevated temperatures~\cite{Song2011,Xu2013}. Extremely narrow ribbons (a few unit cells wide) can be further thinned by broadly focused electron irradiation, eventually leading to atomic chains of carbon~\cite{Chuvilin2009,Jin2009,Song2011}. Similar effects have been observed in hexagonal boron nitride~\cite{Cretu2014} and transition metal dichalcogenides~\cite{Liu2013,Lin2014}. In the last case, it is interesting to note that the material composition was changed from MoS$_{2}$ to approximately  Mo$_{5}$S$_{4}$, since the lighter S atoms are more easily sputtered under the electron beam~\cite{Komsa2012,Liu2013}. Even junctions of nanowires can form from narrow constrictions that were patterned and further thinned by electron irradiation~\cite{Lin2014}. 

\begin{figure}
\includegraphics[width=1\linewidth]{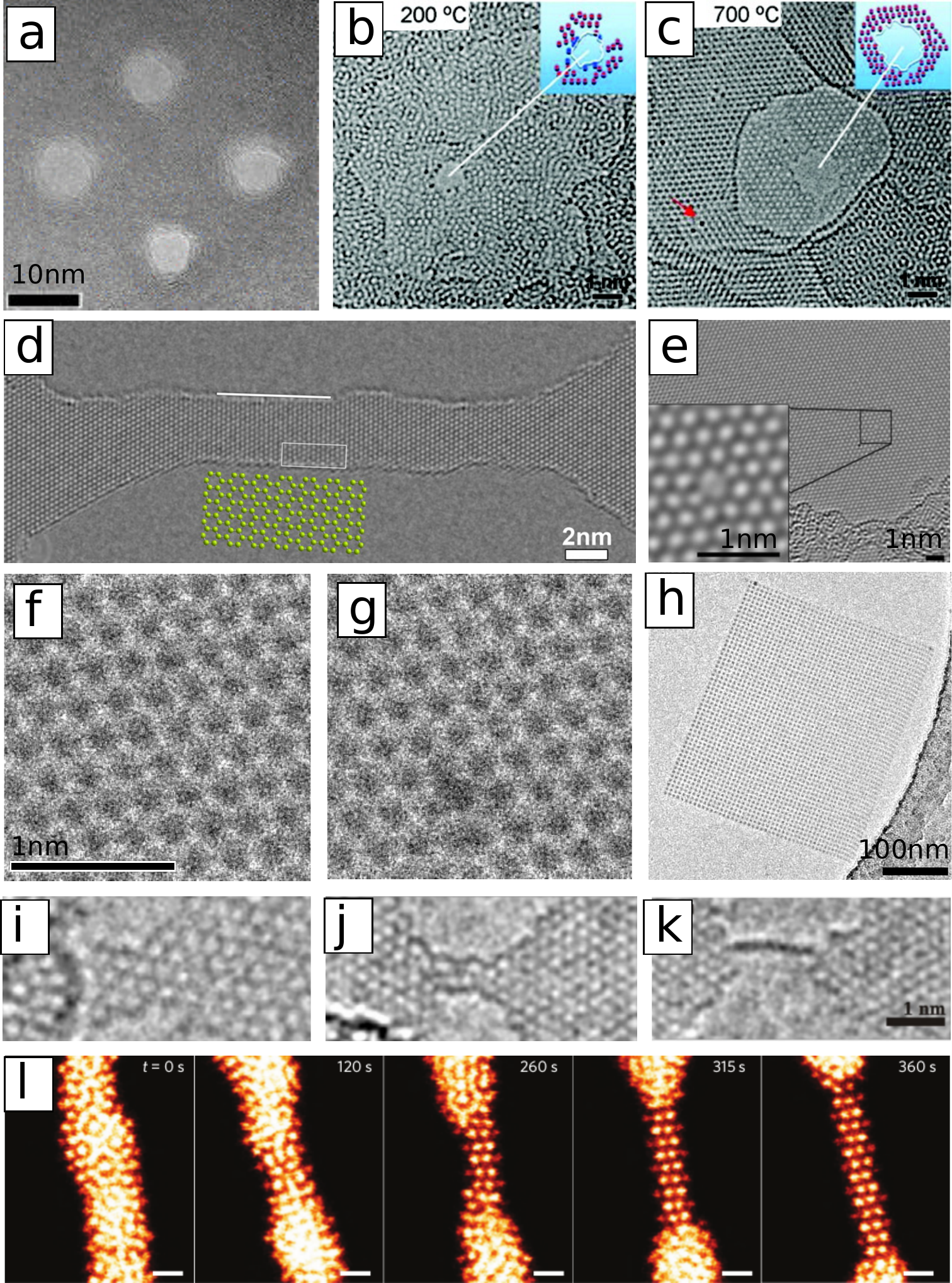}
\caption{{\bf Structural modifications in 2D materials created by local irradiation with electrons.} (a) Pores cut into few-layer graphene by 200-keV irradiation at room temperature~\cite{Fischbein2008}. (b,c) Holes in single-layer graphene formed by 300-keV irradiation at 200\textdegree C (b) and at 700\textdegree C (c)~\cite{Song2011}. (d) Ribbon sculpted from graphene by 300-keV irradiation at 600\textdegree C~\cite{Xu2013}. (e) Divacancy formed by high-intensity 80-keV irradiation at room temperature~\cite{Robertson2012}. (f,g) STEM images of graphene recorded at 90~keV, showing the formation of a vacancy during observation (images from the data of Ref.~\cite{susi_isotope_2016}, available at ~\cite{SusiFigshare2016}). (h) Carbon deposit pattern written onto a graphene membrane~\cite{Meyer2008b}. (i,k) Narrow graphene ribbon converting to a single atomic carbon chain~\cite{Chuvilin2009}. (l) Thin junction of MoSe transforming into a crystalline wire under irradiation~\cite{Lin2014}. (a,h) Reprinted with permission from Refs.~\cite{Fischbein2008,Meyer2008b}, Copyright 2008 AIP Publishing. (b,c,d,e,l) Reprinted with permission from Refs.~\cite{Song2011,Xu2013,Robertson2012,Lin2014} Copyright 2011-2014 American Chemical Society. (i-k) Adapted from Ref.~\cite{Chuvilin2009}.\label{fig:struct}}
\end{figure}

While wires and ribbons are obvious building blocks for future electronic devices, tiny pores in a 2D membrane have potential applications in filtering, energy storage and DNA translocation. Pores down to few-atom vacancies have been fabricated in graphene by the use of a focused e-beam~\cite{Fischbein2008,Song2011,Robertson2012}. Although the formation of large numbers of pores via serial processing with a focused beam is not a viable route for mass production, for research purposes it has the advantage of direct feedback and a high degree of control.

In Fig.~1, we present some examples of e-beam induced modifications of 2D materials. Figs.~1a,b show pores fabricated by focused 200--300-keV irradiation of single-layer graphene at room temperature (a) and at a slightly elevated temperature (b). In the closer view (Fig.~1b), it is clear that the area surrounding the hole is strongly damaged, displaying an amorphous structure. This amorphization is likely due to a significant electron beam current outside the central focused spot. Fig.~1c shows in comparison a pore fabricated by 300-keV irradiation at 700\textdegree C (to stimulate self-healing). Remarkably, the lattice surrounding this pore has remained crystalline. Fig.~1d shows a ribbon that was created using 300-keV irradiation to cut a pattern into graphene at 600\textdegree C. Fig.~1e shows a divacancy generated by focusing an 80-kV TEM e-beam to a $\sim$1~nm spot, thereby generating a very high current density. Interestingly, the formation mechanism of these vacancies is hitherto not fully clarified, since 80-kV STEM imaging has not been observed to lead to vacancy formation in pristine graphene. Fig.~1f,g shows two frames from a sequence of 90-kV STEM images, where a vacancy is formed after a few exposures. In this case, the beam was continuously scanned over a 1~nm area until a defect was created~\cite{susi_isotope_2016,SusiFigshare2016}. An example that is not based on sputtering but rather the deposition of material on graphene is shown in Fig.~1h, where a carbon pattern has been written onto graphene. The carbon source was the residual mobile contamination on the sample or in the microscope column. Clearly, electron beam induced deposition can be used to create structures on graphene with few-nm resolution. Finally, Fig.~1i-l show examples where a narrow junction in graphene (i-k) and a transition metal dichalcogenide (l; MoSe$_2$) was further thinned by broadly focused illumination. The graphene ribbon converted into a single-atomic carbon chain and the MoSe$_2$ ribbon into a MoSe nanowire~\cite{Lin2014}.

\section{Beam-driven dynamics at the level of individual atoms}

Beam-induced changes in a structure during a TEM experiment are usually described as 'radiation damage'~\cite{Egerton2012} and are often connected to the loss of atoms (e.g., via sputtering of atoms at high electron energies). In connection with nanofabrication, however, the structural changes are a desired feature, especially if they can be understood and controlled. At high energies, sputtering is the dominant mechanism and typically holes are formed in the exposed areas, as discussed above. Such ``knock-on'' damage is due to direct momentum transfer from an electron to an atomic nucleus, and can be suppressed by using lower energies~\cite{Egerton2010}. However, effects of beam-induced etching as well as beam-induced material deposition from a precursor (or from contamination) are not strongly dependent on the electron energy. Instead, reducing these beam-activated chemical processes requires clean samples and a clean vacuum environment. 

Aberration correction has enabled the use of lower electron energies for atomic resolution imaging of even light elements. This has been the key to imaging various 1D and 2D materials down to individual atoms~\cite{Suenaga2007,Meyer2008a,Gass2008,Krivanek2010a} without causing damage (e.g., for graphene, at and below about 80~kV~\cite{Meyer2012,susi_isotope_2016}). However, changes in the structure are often not entirely prevented even at low voltages: even if the loss of atoms is suppressed, defects can change their configuration also without sputtering. For example, some impurities, as well as the divacancy in graphene~\cite{kotakoski_imaging_2014}, display beam-induced dynamics at just the right ``speed'' (probability of an event occurring as compared to the electron dose required for recording an image) to follow their transformation in a sequence of STEM or HRTEM images. Fig.~2 shows some examples of such beam-induced dynamics of defects and impurities that can be followed at the atomic level. In contrast to the examples in Fig.~1, these structural changes conserve the number of atoms. 

Changes that involve beam-induced bond rotations in graphene in the absence of impurities have been discussed in many works. The most elementary defect that involves a bond rotation, often called the Stone-Wales defect (Fig.~2a-c), can be formed and annihilated under 80-keV electron irradiation~\cite{Meyer2008a,Kotakoski2011a}. Divacancies in graphene change between three different stable isomers under irradiation due to beam-induced bond rotations~\cite{Kotakoski2011,Kotakoski2011a,Robertson2012,Warner2013,kotakoski2014imaging,Sun2015} (Fig.~2b-d), which are also important in the reconstruction of grain boundaries and more complex defects in graphene~\cite{Huang2011,Warner2012a,Kurasch2012,Lehtinen2013,Warner2013,eder2014journey}.
Impurities in graphene and especially their dynamics under the beam have also been described by various groups. The elements that have been identified as single- or few-atom impurities in graphene include nitrogen~\cite{Meyer2011,Bangert2013,Lin2015}, boron~\cite{Bangert2013}, silicon~\cite{zhou_direct_2012,Zhou2012,lee_direct_2013,ramasse_probing_2013,susi_siliconcarbon_2014,Wang2014} (Fig.~2e-f), phosphorus~\cite{Susi172DM}, and iron~\cite{Robertson2013,Zhao2014}, whereas several other metals catalyze a destruction of graphene under the e-beam~\cite{Ramasse2012}. Atoms at the edge of a graphene sheet remain highly mobile under 60--80-keV irradiation~\cite{Girit2009,Chuvilin2009,Suenaga2011}, including impurity atoms (Fig. 1g,h)~\cite{Wang2014,zhao_direct_2014}. However, dynamics at the edge do not usually conserve the number of atoms over longer image sequences~\cite{Wang2014}, apart from individual changes in configuration such as the migration of atoms along an edge or the transformation between reconstructed and non-reconstructed edge configurations~\cite{Chuvilin2009}.

\begin{figure}
\centering\includegraphics[width=0.8\linewidth]{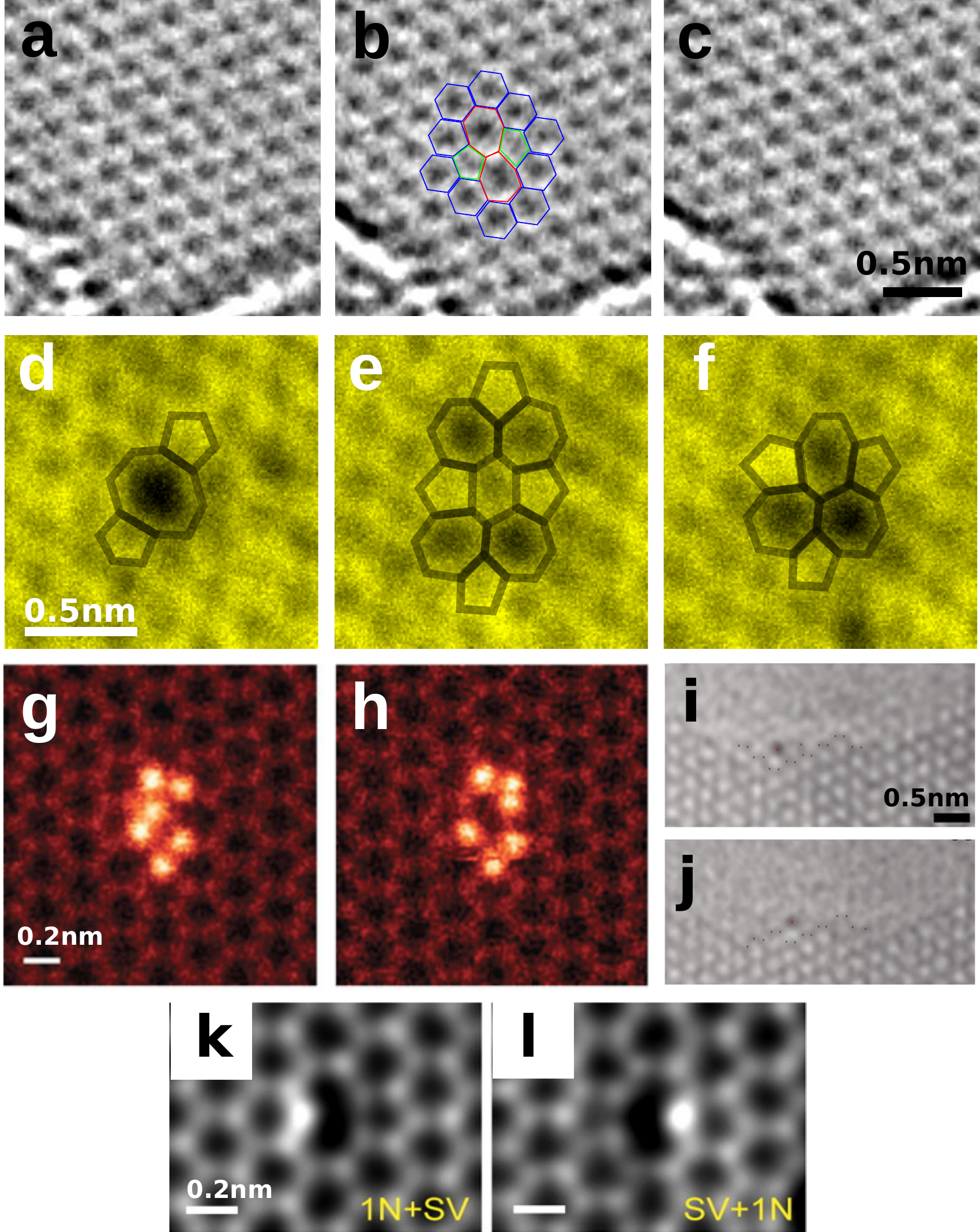}
\caption{{\bf Examples of atom-number-conserving beam-driven dynamics.} (a-c) Stone-Wales defect that is formed and annihilated under 80-keV electron irradiation~\cite{Meyer2008a}. (d-f) Divacancy converting between three stable configurations under the beam~\cite{kotakoski2014imaging}. (g-h) Cluster of Si impurities in a graphene pore switching between different configurations~\cite{lee_direct_2013}. (i-j) A heavier atom bound at the edge of graphene, translated along the edge\cite{zhao_direct_2014}. (k-l) Pyridinic nitrogen--vacancy defect switching between two configurations \cite{lin_pyridinicN_2015}. (a,b,c,k,l) Reprinted with permission from Refs.~\cite{Meyer2008a,lin_pyridinicN_2015}, Copyright 2008 and 2015 American Chemical Society. (g,h) Reprinted with permission from Ref.~\cite{lee_direct_2013}, Copyright Nature Publishing Group. (i,j) Adapted from Ref.~\cite{zhao_direct_2014}, Copyright 2014 National Academy of Sciences.}
\end{figure}

Atom-number-conserving, and potentially reversible displacements of impurities under the influence of a local probe are an important prerequisite for creating structures at the atomic level. Another one is that the atomic displacements can be steered in a desired direction. One of the impurities, silicon (Si), appears to fulfill these prerequisites. In addition, it is an ubiquitous contamination, frequently found incorporated into the graphene lattice as isolated substitutions. While the origin of the Si contamination remains unclear, we have found Si in samples of exfoliated graphene, graphene made by chemical vapor deposition, and in samples reduced from graphene oxide. Irrespective of their origin, Si impurities have provided a playground for single-atom dynamics as well as for the spectroscopic analysis of different bonding configurations~\cite{Ramasse2013,zhou_direct_2012,Zhou2012,lee_direct_2013}. Fig.~3a reproduces images from Ref.~\cite{susi_siliconcarbon_2014}, where the displacement of Si atoms under the e-beam was analyzed and the measured cross sections for the displacement were compared to simulations. Importantly, the analysis revealed that the displacement of Si was triggered by an electron knock-on event on a {\em carbon atom next to the Si} instead of the Si atom itself (Fig~3b,c). This is the key to steering the displacement in a particular direction. Hence, it was proposed~\cite{susi_siliconcarbon_2014} that focusing the e-beam on the carbon atom next to the Si would result in an exchange of places between the Si and C, as illustrated in Fig.~3d. However, until now this has not been demonstrated in a controlled experiment.

\begin{figure}[t]
\centering\includegraphics[scale=0.4]{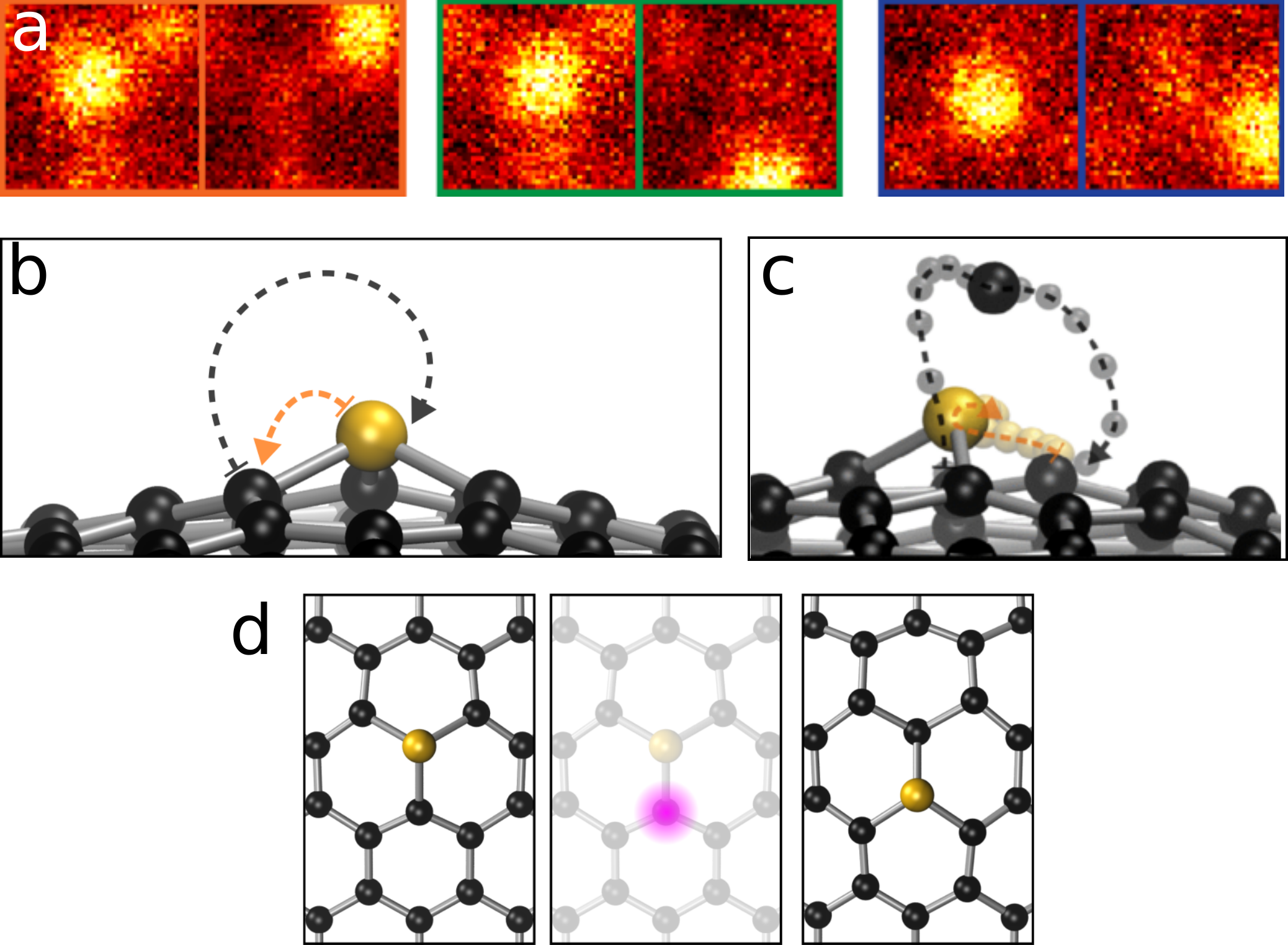}
\caption{{\bf Lattice-site jump of a Si atom in graphene.} (a) Displacements of individual, three-coordinated Si atoms as observed in Ref.~\cite{susi_siliconcarbon_2014}. (b) Proposed mechanism for the movement: a C atom is hit by a fast electron, goes out of the plane and exchanges places with the Si atom. (c) Actual trajectory from an atomistic simulation~\cite{susi_siliconcarbon_2014}. (d) An illustration
of beam-controlled inversion of the Si--C bond in graphene: The location of the silicon substitution (yellow sphere) is first identified by scanning an image. Then, the beam is placed on top of the carbon atom in the lattice site where the silicon is desired to be moved. Each electron has a small probability to pass close enough to the carbon nucleus so that it transfers significant momentum to it, sending it on an out-of-plane path. In the end, the location of the Si atom is changed by one lattice site without the loss of atoms. Adapted from~\cite{susi_siliconcarbon_2014}.\label{Fig1}}
\end{figure}

\section{Single-atom manipulation}

Until now, the most remarkable successes in the manipulation of individual atoms have been achieved with scanning tunneling microscopy (STM), where atoms can be picked up from a shallow local potential energy minimum by voltage pulses from the atomically sharp conducting tip and placed at the location of another energy minimum~\cite{crommie_confinement_1993}. As a recent display of the power of this method, Kalff and co-workers demonstrated~\cite{kalff_kilobyte_2016} a re-writable atomic-scale memory array with a capacity of a kilobyte (8000 bits). Their device reached an outstanding areal density of 502 terabits per square inch (ca. 0.778 per nm$^{2}$), but due to the nature of the data storage medium---vacancies in a monolayer of chlorine on a Cu(100) surface---the data corrupts at temperatures above 77 K. It is easy to predict, at least within a simplistic model, that the manipulation of atoms on a surface by STM will be limited to low temperatures and specific combinations of atoms: any method that moves atoms from one location to another must obey a hierarchical relation of energies:
\[ k_{B}T\ll E_{barrier}<E_{tool},\]
with thermal energy $k_{B}T$, barrier for atom relocation $E_{barrier}$, and interaction with the atom sized tool $E_{tool}$ ($k_B$ being the Boltzmann constant and $T$ the absolute temperature). For STM-based manipulation, the interaction between the tip and the surface atom can be estimated to be on the order of 1~eV (typical voltages on the STM tip), while for atoms on a surface, the barrier for their relocation is an order of magnitude lower~\cite{kalff_kilobyte_2016}; hence, low temperatures are required for the atoms to stay where they are. In contrast, atomic lattice impurities in a 2D material have migration barriers of several eV, which makes them stable at room temperature. In addition, the energy that can be transferred from a focused electron beam to an atom in a knock-on process is high enough to overcome this barrier. Below, we show initial results that demonstrate the controlled movement of Si in a graphene lattice as we proposed in Ref.~\cite{susi_siliconcarbon_2014}. 

\subsection{Methods and materials}

Our experiments were conducted using a Nion UltraSTEM100 scanning transmission electron microscope, operated at 60~kV in near-ultrahigh vacuum ($2\times10^{-7}$~Pa). This instrument provides a sufficient stability in the sample stage and electronics to allow (predominantly) irradiating just one atom at a time over time spans of several tens of seconds. The typical beam current of the instrument under our experimental conditions is 50~pA. The beam convergence semiangle was 30 mrad and the semi-angular range of the medium-angle annular-dark-field (MAADF) detector was $60-200$~mrad. As samples we used commercial graphene grown by chemical vapor deposition and transferred onto TEM grids (Graphenea {\textregistered } R 2/4), with known ubiquitous silicon contamination in the lattice~\cite{zhou_direct_2012,ramasse_probing_2013,susi_siliconcarbon_2014}. 

\subsection{Moving individual Si atoms in graphene}

\begin{figure}
\centering\includegraphics[scale=1.4]{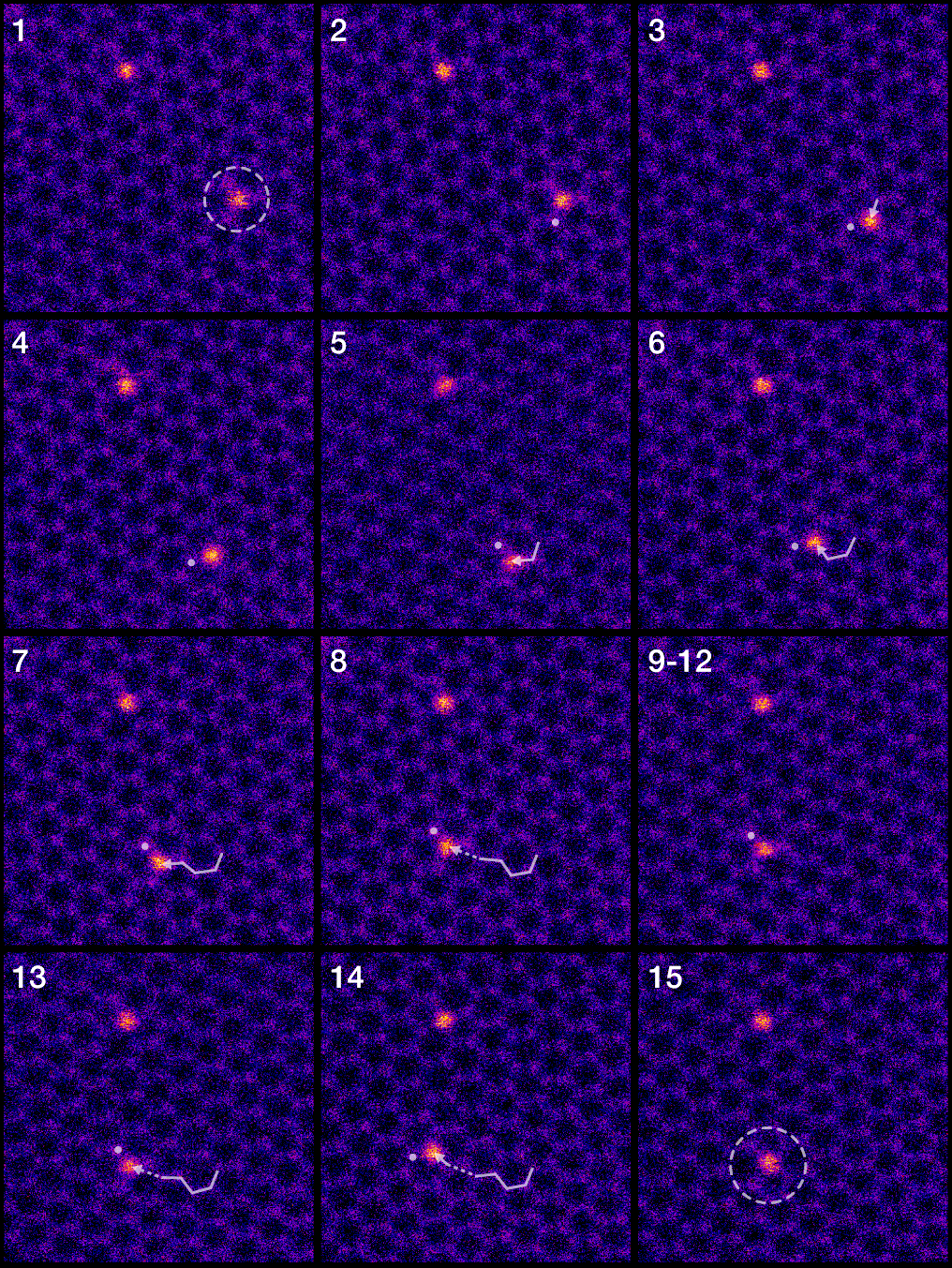} \caption{{\bf An example STEM image series of Si atom manipulation in graphene} (MAADF detector, 60~kV, 8 $\mu$s pixel dwell time). In the first frame, the three-coordinated Si impurity selected for manipulation is denoted by the dashed circle. In the following frames, the solid dots denote the location of the parked beam, the solid arrows the cumulative single-site jumps, and dashed arrows jumps over more than one site. The 512$\times$512~px frames have been cropped from the original 1024$\times$1024 px fields of view to manually correct for drift (less than 2 {\AA} over the 6 min series), and colored with the ImageJ lookup table ``Fire''.\label{Fig2}}
\end{figure}

\begin{figure}
\centering\includegraphics[scale=1.25]{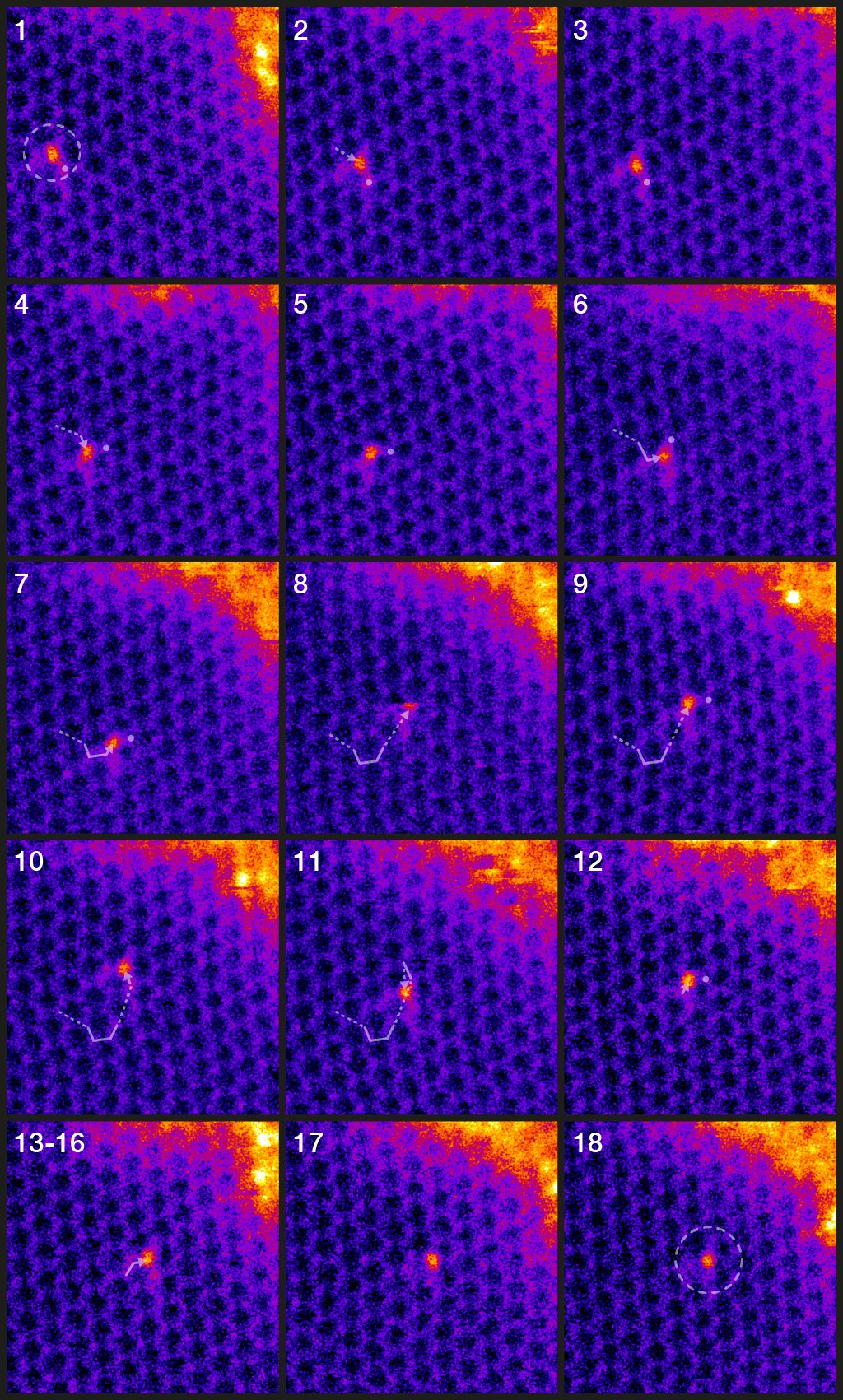} \caption{{\bf Another STEM image series of Si atom manipulation in graphene} (capture parameters as in Fig.~\ref{Fig2}, but a 1 px Gaussian blur applied to reduce noise).\label{Fig3}}
\end{figure}

First, we survey the sample and find a clean area of monolayer graphene with one or more embedded Si atoms, which appear brighter than the carbon atoms due their higher scattering contrast~\cite{krivanek_Z-contrast_2010}. After choosing a Si for our manipulation and capturing a field of view over the area, we stop the scan and park the beam on top of a selected C neighbor to the Si. Using a Python plugin coded for the Nion Swift control software, we set an irradiation time (typically 15~s) after which the microscope automatically captures another frame and again parks the beam onto the cursor location. If the Si atom has moved or the sample has slightly drifted, this location is manually altered to ensure that the correct atom is being predominantly irradiated. This procedure is then repeated iteratively to move the atom through the lattice. An example of the single-atom manipulation experiments is shown as Figure~\ref{Fig2}, and another one as Figure~\ref{Fig3}.

In both cases, we were able to control the movement of the Si by several lattice sites, with occasional jumps over more than one site (Fig.~\ref{Fig2} frame 8, Fig.~\ref{Fig3} frames 4, 8 and 11) or into the wrong direction (Fig.~\ref{Fig3} frames 8, 10 and 11). Both experiments were terminated after a C neighbor of the Si was ejected, resulting in the four-coordinated planar configuration~\cite{ramasse_probing_2013}, which cannot be moved further. In our original study on this topic~\cite{susi_siliconcarbon_2014}, these almost invariably healed back into the lower-energy three-coordinated substitution within a minute or so, but in these experiments this did not happen. A certain amount of mobile carbon adatoms would thus be beneficial, but this seems sample and/or location dependent.

\subsection{Discussion}

Unfortunately, the degree of control over the atoms' movement is (so far) not as good as in the case of the STM. As we see it, there are three main hurdles to overcome: 1) when placing the beam on the C atom towards which we want the Si to move, we are blind to changes in the structure; 2) all our trials were terminated when one C neighbor to Si was knocked out, resulting in the four-coordinated configuration which cannot be moved further; and 3) the Si sometimes jumps in either the wrong direction, or over several lattice sites between the captured frames.

The first issue is easy to mitigate with improvements in hardware-software integration: by reading out the scattering signal in real time, the spot irradiation could be automatically terminated once an increase corresponding to the appearance of the heavier atom is observed. Alternatively, a small subscan window centered on the C atom could be used, and the integrated signal or user intervention used as a terminating signal. The second issue might be mitigated by lowering the acceleration voltage: since the threshold for C displacement is about 2~eV higher than for the bond inversion~\cite{susi_siliconcarbon_2014}, a lower electron energy would suppress damage while still retaining a finite probability for the jumps. Since atomic resolution can be retained down to 40~kV in the Nion UltraSTEM~\cite{Dellby_40kV}, there is room to optimize the voltage. The third issue most likely results from multiple impacts during irradiation, which would be mitigated by the solution to the first issue, or from the probe tails during the scan itself, which possibly might be mitigated by optimizing the parameters of the probe-forming electron optics.

Separate from the manipulation itself, sample quality presents a significant hurdle. Monolayer graphene is notoriously dirty on the atomic level, and finding clean areas, especially containing heteroatoms, is frustratingly difficult. Even when such areas are found, often the contamination gathers under the beam and obscures the surface. Finally, a sufficient quantity of heteroatoms need to be introduced into the lattice in the first place, but without causing significant structural damage.

To address the first two sample issues, in-situ cleaning methods or heating the sample during observation seem to be possible solutions. For the latter, ion implantation seems to be a promising technique, but likely requires carefully selected and narrowly distributed ion energies for success~\cite{Ahlgren2011}, as well as transferring the sample in vacuum after implantation.

\section{Conclusions}

Focused electron beams are a versatile tool for shaping low-dimensional materials at a scale down to single atoms. Besides manipulation, the direct feedback that is provided by simultaneous imaging is of key importance. Moreover, the possible variability in electron energy (while maintaining resolution) that is afforded by modern aberration-corrected instruments provides a handle for controlling sputtering in comparison to other beam-driven structural changes. 

We have demonstrated the first steps towards the atomic precision manipulation of silicon impurities embedded in the graphene lattice using the scanning transmission electron microscope. At this very early stage, although several hurdles remain in making our method practical, we do not see any fundamental physical roadblocks to significant progress. Apart from scanning tunneling microscopy, which is limited to weakly bound surface atoms at low temperatures, no other technique we are aware of can even in principle achieve such control. Since the properties of materials are determined by their chemical structure, single-atom manipulation can be considered to be the ultimate goal for both materials science and engineering.

\section*{Acknowledgments}

T.S. acknowledges funding from the Austrian Science Fund (FWF) via project P~28322-N36, and the computational resources of the Vienna Scientific Cluster. J.C.M. acknowledges funding by the European Research Council Grant No. 336453-PICOMAT. J.K. acknowledges funding from the Wiener \mbox{Wissenschafts-,} Forschungs- und Technologiefonds (WWTF)
via project MA14-009.

\section*{References}

\bibliographystyle{ieeetr}
\bibliography{ultram_bibfile,library}

\end{document}